\documentclass[twocolumn,showpacs,preprintnumbers,amsmath,amssymb]{revtex4-1}

\usepackage{graphicx}
\usepackage{dcolumn}
\usepackage{bm}

\begin{document}

\title{Non-Entangling Channels for Multiple Collisions of Quantum Wave Packets}

\author{Walter Hahn}
\email{W.Hahn@yandex.ru}
\affiliation{Institute for Theoretical Physics, University of Heidelberg, Philosophenweg 19, 69120 Heidelberg, Germany}
\author{Boris V. Fine}
\email{B.Fine@thphys.uni-heidelberg.de}
\affiliation{Institute for Theoretical Physics, University of Heidelberg, Philosophenweg 19, 69120 Heidelberg, Germany}

\date{April 27, 2011}

\begin{abstract}
We consider  multiple collisions of quantum wave packets in one dimension. The system under investigation consists of an impenetrable wall and of two hard-core particles with very different masses. The lighter particle bounces between the heavier one and the wall. Both particles are initially represented by narrow Gaussian wave packets. A complete analytical solution of this problem is presented on the basis of a new method. The idea of the method is to decompose the two-particle wave function into a continuous superposition of terms (channels), such that the multiple collisions within each channel do not lead to subsequent entanglement between the two particles. For each channel, the time evolution of the two-particle wave function is completely determined by the motion of the corresponding classical point-like particles; therefore the whole quantum problem is reduced to a classical calculation.  The calculation based on the above method reveals the following unexpected result: The entanglement between the two particles first increases with time due to the collisions, but then it begins to decrease, disappearing completely when the light particle becomes too slow to catch up with the heavy one.
\end{abstract}


\maketitle

In quantum mechanics, particles are often represented by wave packets. According to the Schrödinger equation, the wave packet of a free particle typically spreads without a limit. This spread was of concern in the early days of quantum mechanics for Schrödinger \nolinebreak himself \nolinebreak \cite{schroedharm}, because it had never been observed for macroscopic systems. He was able to identify the non-spreading coherent wave packets only for the harmonic potential. In the free particle case, the spread of the wave packets and the accompanying winding of the quantum phase significantly complicate practical calculations, in particular, in the treatment of multiple collisions between particles. Typically, each collision between two quantum wave packets leads to the entanglement between them \cite{schulmanwall,freyberger,schulman,law,translentang,schmujan,harshhutton,kamleitner}. As a result, the wave function of the system after many collisions becomes increasingly intractable.

The studies of colliding wave packets are often motivated either by the agenda of the controlled generation of entanglement \cite{freyberger,law,translentang,schmujan,harshhutton} or by the studies of quantum decoherence \cite{schulmanwall,schulman,kamleitner}. In the both contexts, the treatment of multiple collisions between quantum wave packets is of significant interest.
In the former context, it was investigated for two cold atoms in a harmonic trap in Ref.\cite{freyberger}.

\begin{figure}[t]
		\centering
		\scalebox{0.19}{\includegraphics{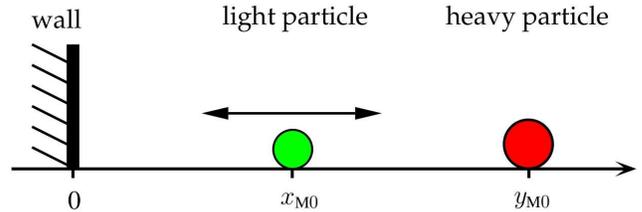}}
		\caption{(Color online) The system considered: the light particle bouncing between the heavy particle and the wall. }
		\label{pic:final}
\end{figure}

In this article, we consider the problem of a light particle bouncing between a heavy particle and a wall, shown in Fig.~\ref{pic:final}, and propose a new method for dealing with multiple scattering of quantum wave packets. This method is based on the observation that there exist special initial conditions which do not lead to entanglement between two colliding Gaussian wave packets. We refer to states with these conditions as ``non-entangling channels''. The method consists of decomposing the initial wave function of the system as a superposition of non-entangling channels. The calculation for each channel becomes largely classical, even though the complete solution retains all the usual quantum features. For the two-particle problem considered, we report a remarkable result: Even though the wave packets of the two particles become entangled after initial collisions, the entanglement begins to decrease later and eventually disappears completely.

The setting of our problem is shown in Fig.~\ref{pic:final}. We use the variable $x$ for the coordinate and for all subscripts referring to the light particle. Correspondingly, the variable $y$ refers to the heavy particle. The two particles have very different masses $m_x$ and $m_y$ such that $\epsilon=\sqrt{m_x/m_y}\ll1$.
We assume the particles to be non-relativistic. They interact via the hard-core potential. The wall is represented by an infinitely high potential step. The Hamiltonian of the problem is 
	\begin{equation}
	\hat{H}=\frac{\hat{p}_x^2}{2m_x}+\frac{\hat{p}_y^2}{2m_y}+S_1\Theta(x-y)+S_2\Theta(-x),\label{eqn:ham}
	\end{equation}                                                
where $\hat{p}_x$ and $\hat{p}_y$ are momentum operators and $\Theta(x)$ is the Heaviside-step-function. We only consider the ``hard-core'' limits $S_1\rightarrow\infty$ and $S_2\rightarrow\infty$, therefore the light particle always remains localized between the heavy particle and the wall. 

Initially, the light particle is moving while the heavy particle is at rest. We assume a nearly factorized initial state of two Gaussian wave packets (with $\hbar=1$):
	\begin{eqnarray}
	\psi_0(x,y)\!\sim\!\exp\!\left\{\!-\frac{[x-x_{\hbox{\scriptsize M}0}]^2}{2\sigma_{0x}^2}\!+\!ip_{x0}x\!-\!\frac{[y-y_{\hbox{\scriptsize M}0}]^2}{2\sigma_{0y}^2}\!\right\}\!,\label{eqn:initial}
	\end{eqnarray}
where $0<x_{\hbox{\scriptsize M}0}<y_{\hbox{\scriptsize M}0}$ are the maxima of the Gaussians, $p_{x0}>0$ the initial momentum of the  light particle and $\sigma_{0x}$ and $\sigma_{0y}$ the widths of the respective wave packets, which are assumed to be much smaller than any of the three distances: $x_{\hbox{\scriptsize M}0}$, $y_{\hbox{\scriptsize M}0}$ and $y_{\hbox{\scriptsize M}0} - x_{\hbox{\scriptsize M}0}$. The sign ``$\sim$'' in Eq.(\ref{eqn:initial}) means that the wave function needs to be slightly modified to remove the overlap of the exponentially small tails of the particles' wave functions and the penetration of these tails into the wall. A possible form this modification is given in \cite{elsupp}. 
We further consider only sufficiently large values of $p_{x0}$, such that the widths of both wave packets remain much smaller than $y_{\hbox{\scriptsize M}0}$ until the collisions between the two particles stop (see below).
Therefore, we use the approximation that every collision involves only two constituents: either the two particles or the light particle and the wall.

Let us first recall that a Gaussian wave packet, initially with the maximum at $x_{\hbox{\scriptsize M}0}$, the width $\sigma_{0x}$ and momentum $p_{x0}$ evolves according to a free-particle Schrödinger equation as
	\begin{equation}
	\phi(x;t)=N(t)\exp\left\{-\frac{[x-x_c(t)]^2}{2\beta_x(t)^2}+ip_{x0} x\right\}\label{eqn:evolve},
	\end{equation}
where $\beta_x(t)^2=\sigma_{0x}^2\left(1+\frac{it}{m_x \sigma_{0x}^2}\right)$ is the width parameter, $x_c(t)=x_{\hbox{\scriptsize M}0}+\frac{p_{x0}}{m_x}t$ is the time-dependent position of the maximum and $N(t)$ is the normalization factor. The probability distribution of positions $|\phi(x;t)|^2$ remains Gaussian. Its width grows with time as $|\beta_x(t)^2|$.  We also introduce the analogous width parameter for the heavy particle $\beta_y(t)^2=\nolinebreak\sigma_{0y}^2\left(1+\frac{it}{m_y \sigma_{0y}^2}\right)$. The above-mentioned narrow-wave-packet assumption implies that \begin{equation}
|\beta_y(t)|, |\beta_x(t)| \ll y_{\hbox{\scriptsize M}0}.
\label{ballistic}
\end{equation}

We use the description of the reflection of the light particle from the wall given in Ref.\cite{schulmanwall}. This description incorporates the wall by considering the antisymmetric superposition of the free-particle Gaussian and its mirror image with respect to the position of the wall, i.e. $\phi(x;t) - \phi(-x;t)$. Far away from the wall $\phi(x;t)$ dominates before the collision, while $\phi(-x;t)$ dominates after the collision. Hence, the description of the particle-wall reflection amounts to the transformation $x\rightarrow -x$ in the free-particle Gaussian \eqref{eqn:evolve} \cite{elsupp}.

The single collision of two wave packets via a hard-core potential in one dimension has been described in Refs.\cite{schulman,harshhutton,schmujan}. If the Hamiltonian of the two-particle interaction is rewritten using the center-of-mass coordinate $R = (m_x x + m_y y)/(m_x + m_y)$ and the relative coordinate $r=x-y$, the problem separates into a free-particle motion for variable $R$ and the wall reflection problem for variable $r$. By analogy with the particle-wall collision, the two-particle collision can be described by the transformation $r\rightarrow -r$ with the subsequent transformation back to the variables $(x,y)$ \cite{elsupp}. If the initial wave function has the form \eqref{eqn:initial}, then, after one collision between the two particles, the wave function becomes
	\begin{equation}
	\psi_1(x,y;t)\sim\exp\Big\{A_1^{xx}x^2\!+\!A_1^{yy}y^2\!+\!A_1^{xy}xy\!+\!B_1^xx\!+\!B_1^yy\Big\},\nonumber
	\end{equation}
where all coefficients are functions of time and other parameters of the problem. In general, this wave function is entangled in terms of variables $x$ and $y$ because of the term $A_1^{xy}xy$ appearing in the exponent. However, this term vanishes\cite{schulmanwall,schulman,schmujan,harshhutton,kamleitner}, i.e. $A_1^{xy}=0$, when\footnote{Another case of vanishing entanglement is $m_x=m_y$.}
	\begin{equation}\label{eqn:btemp}
	m_x\sigma_{0x}^2=m_y\sigma_{0y}^2 .
	\end{equation}
In this case, not only the entanglement vanishes but also the width parameters $\beta_x(t)$ and $\beta_y(t)$ continue evolving as in the respective free-particle cases, i.e. after the collision the wave function \nolinebreak is
	\begin{eqnarray}
	\psi_1(x,y;t)&\sim&\exp\left\{-\frac{\left[x-x_{\hbox{\scriptsize M}}(t)\right]^2}{2\beta_x(t)^2}+ip_{x1}x\right\}\label{eqn:wft}\\
	&&\times\exp\left\{-\frac{\left[y-y_{\hbox{\scriptsize M}}(t)\right]^2}{2\beta_y(t)^2}+ip_{y1}y\right\},\nonumber
	\end{eqnarray}
where $p_{x1}$ and $p_{y1}$ are the momenta and $x_{\hbox{\scriptsize M}}(t)$, $y_{\hbox{\scriptsize M}}(t)$ are the maxima of the wave packets moving as the coordinates of two reflected classical particles. The relation (\ref{eqn:btemp}) implies that $m_x\beta_x(t)^2=m_y\beta_y(t)^2$ holds both before and after the collision. It also implies that the kinetic energies stored in the momentum distributions of each of the two particles in the rest frames of their respective wave packets are equal to each other -- the situation somewhat reminiscent of the thermal equilibrium.

We observe further: Since the evolutions of the parameters $\beta_x(t)$ and $\beta_y(t)$ remain unaffected after a two-particle collision in the case of the condition (\ref{eqn:btemp}) and the same is true for the particle-wall collision, the ansatz of the form (\ref{eqn:wft}) remains valid after multiple collisions of the light particle with the wall and with the heavy particle. The evolutions of the centers of the Gaussian wave packets $x_{\hbox{\scriptsize M}}(t)$ and $y_{\hbox{\scriptsize M}}(t)$ and their respective momenta can be calculated by identifying the wave packet's maxima with the coordinates of classical point-like particles. This is what we call ``non-entangling channels''.

We now turn to the general case when the wave packets do not obey the relation~\eqref{eqn:btemp}. We limit ourselves to the case
	\begin{equation}
	m_x\sigma_{0x}^2<m_y\sigma_{0y}^2.
	\label{eqn:zerlegen}
	\end{equation}
(The opposite inequality affords qualitatively the same treatment.) The inequality \eqref{eqn:zerlegen} is equivalent to the statement that the initial wave packet for the heavy particle is too broad to obey the relation \eqref{eqn:btemp}. Central to our method is the representation of the initial wave packet for the heavy particle as a superposition of Gaussian wave packets with different maxima but the same width, which is smaller than the initial one $\sigma_{0y}$ such that it obeys Eq.\eqref{eqn:btemp}. That is, each of the wave packets in this decomposition together with the wave packet for the light particle describe a non-entangling channel. In this way, we reduce the whole quantum problem to the classical calculation of the probability distribution for the non-entangling channels.

We divide $\sigma_{0y}^2$ into two parts
	\begin{equation}
	\sigma_{0y}^2=\Delta\sigma_{y,0}^2+\sigma_{yT}^2,\label{eqn:behave}
	\end{equation}
where $\sigma_{yT}^2$ is defined by the condition $m_x\sigma_{0x}^2=m_y\sigma_{yT}^2$ [cf. Eq.\eqref{eqn:btemp}], and rewrite the initial wave function in Eq.\eqref{eqn:initial} using the mathematical fact that the convolution of two Gaussian functions is a Gaussian function again:
	\begin{eqnarray}
	&\psi_0&(x,y)\sim\exp\left\{-\frac{[x-x_{\hbox{\scriptsize M}0}]^2}{2\sigma_{0x}^2}+ip_{x0}x\right\}\label{eqn:herehere}\\
	&&\times\int_{-\infty}^\infty dy_{\hbox{\scriptsize m}}\exp\!\left\{\!-\frac{\left[y_{\hbox{\scriptsize m}}-y_{\hbox{\scriptsize M}0}\right]^2}{2\Delta\sigma_{y,0}^2}\!\right\}\exp\!\left\{\!-\frac{[y-y_{\hbox{\scriptsize m}}]^2}{2\sigma_{yT}^2}\!\right\}\nonumber\!.
	\end{eqnarray}
The first Gaussian function inside the integral can be interpreted as the distribution of the maxima $y_{\hbox{\scriptsize m}}$ for the heavy particle in different non-entangling channels. We also introduce the variable $x_{\hbox{\scriptsize m}}$ for the maxima of the light particle for each non-entangling channel. According to Eq.(\ref{eqn:herehere}), all $x_{\hbox{\scriptsize m}}$ initially coincide with $x_{\hbox{\scriptsize M}0}$. 

The problem is now reduced to calculating the time evolution of the positions $x_{\hbox{\scriptsize m}}$ and $y_{\hbox{\scriptsize m}}$ of classical particles and then obtaining the probability distribution of these positions, which we denote as $P_{\hbox{\scriptsize xy},n}(x_{\hbox{\scriptsize m}},y_{\hbox{\scriptsize m}};t)$.  Here index $n$ represents the number of collisions between the two particles that occurred before time $t$ in each non-entangling channel.
Given our assumption of sufficiently narrow wave packets, we only consider the moments of time between the collisions when particles in each non-entangling channel have experienced the same number of collisions and, otherwise, are sufficiently far from each other and from the wall.  Therefore, assigning the same index $n$ to all non-entangling channels at a given moment of time is justified. Since the momenta of the particles do not change between the collisions, their respective values $p_{xn}$ and $p_{yn}$ only depend on the number of preceding collisions $n$ but not on the channel and not on time.
Below we also use marginal probability distributions
$P_{\hbox{\scriptsize x},n}(x_{\hbox{\scriptsize m}};t) \equiv \int_{-\infty}^{\infty} dy_m P_{\hbox{\scriptsize xy},n}(x_{\hbox{\scriptsize m}},y_{\hbox{\scriptsize m}};t)$ and $P_{\hbox{\scriptsize y},n}(y_{\hbox{\scriptsize m}};t) \equiv \int_{-\infty}^{\infty} dx_m P_{\hbox{\scriptsize xy},n}(x_{\hbox{\scriptsize m}},y_{\hbox{\scriptsize m}};t)$.

The time evolution of the wave function of the system can now be represented as
	\begin{eqnarray}\label{eqn:nunhab}
	&\psi_n&(x,y;t)\sim\int_{-\infty}^\infty\int_{-\infty}^\infty dx_{\hbox{\scriptsize m}}dy_{\hbox{\scriptsize m}}\ P_{\hbox{\scriptsize xy},n}(x_{\hbox{\scriptsize m}},y_{\hbox{\scriptsize m}};t)\\
	&&\exp\!\left\{\!-\frac{[x-x_{\hbox{\scriptsize m}}]^2}{2\beta_x^2(t)}\!\pm\!i p_{xn}x\!\right\}\exp\!\left\{\!-\frac{[y-y_{\hbox{\scriptsize m}}]^2}{2\epsilon^2\beta_x^2(t)}\!+\!ip_{yn}y\!\right\}\!,\nonumber
	\end{eqnarray}
where sign $\pm$ implies $+$, when the light particle moves away from the wall, and $-$, when it moves towards the \nolinebreak wall.

\begin{figure}[b]
	\centering
	\scalebox{0.19}{\includegraphics{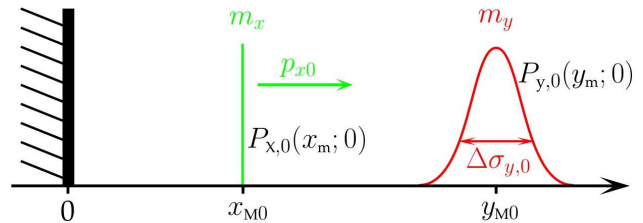}}
	\caption{(Color online) Classical probability distributions: green - $P_{\hbox{\scriptsize x},0}(x_{\hbox{\scriptsize m}};0)$; red - $P_{\hbox{\scriptsize y},0}(y_{\hbox{\scriptsize m}};0)$. }
	\label{pic:classprobinit}
\end{figure}

According to Eq.(\ref{eqn:herehere}), the initial probability distribution for the underlying classical problem is 
\begin{eqnarray}\label{eqn:initcond}
	P_{\hbox{\scriptsize xy},0}(x_{\hbox{\scriptsize m}},y_{\hbox{\scriptsize m}};0)=N_0 \ &&\exp\left\{-\frac{\left[y_{\hbox{\scriptsize m}}-y_{\hbox{\scriptsize M}0}\right]^2}{2\Delta\sigma_{y,0}^2}\right\}\\
	&&\times\delta\left(x_{\hbox{\scriptsize m}}-x_{\hbox{\scriptsize M}0}\right),\nonumber
	\end{eqnarray}
where $N_0$ is the normalization constant. The corresponding marginal probability distributions $P_{\hbox{\scriptsize x},0}(x_{\hbox{\scriptsize m}};0)$ and $P_{\hbox{\scriptsize y},0}(y_{\hbox{\scriptsize m}};0)$ are illustrated in Fig.~\ref{pic:classprobinit}: The light particle has a definite initial position $x_{\hbox{\scriptsize M}0}$ and the momentum $p_{x0}$, while the heavy particle has a Gaussian distribution of possible initial positions $y_{\hbox{\scriptsize m}}$ and no momentum. The subsequent evolution of $P_{\hbox{\scriptsize xy},n}(x_{\hbox{\scriptsize m}},y_{\hbox{\scriptsize m}};t)$ is obtained below in the leading order in the small parameter $\epsilon = \sqrt{m_x/m_y}$.

In each two-particle collision, the light particle transfers momentum to the heavy particle. Eventually, the collisions stop when the light particle becomes too slow to catch up with the heavy particle. The last collision will have index $ n_{\hbox{\scriptsize max}}\approx\frac{\pi}{4\epsilon}$ \cite{elsupp}.

The evolution of the probability distribution $P_{\hbox{\scriptsize xy},n}(x_{\hbox{\scriptsize m}},y_{\hbox{\scriptsize m}};t)$ between the collisions, i.e. for the fixed value of $n$, just amounts to the shift of the values of  $x_{\hbox{\scriptsize m}}$ and $y_{\hbox{\scriptsize m}}$ associated with the motion with respective velocities $p_{xn}/m_x$ and $p_{yn}/m_y$.
Therefore, the shape of $P_{\hbox{\scriptsize xy},n}(x_{\hbox{\scriptsize m}},y_{\hbox{\scriptsize m}};t)$ only changes when $n$ changes.

Two trends in the evolution of the probability distribution $P_{\hbox{\scriptsize xy},n}(x_{\hbox{\scriptsize m}},y_{\hbox{\scriptsize m}};t)$ as a function of $n$ can be anticipated, namely, the initial increase in the spread of the possible values of $x_m$ and the initial decrease in the spread of the possible values of $y_m$. Both trends originate from the time delay between collisions for different initial positions of the heavy particle $y_{m0}$: The larger $y_{m0}$, the later the first and the subsequent collisions occur. The effect of each such time delay is rather small, but it accumulates after many collisions.
The resulting evolution of the particle coordinates becomes \cite{elsupp}:
	\begin{eqnarray}
	x_{\hbox{\scriptsize m}}(y_{\hbox{\scriptsize m}0};t)&=&x_{\hbox{\scriptsize M}}(t)+\frac{\sin(2\epsilon n)}{\epsilon}(y_{\hbox{\scriptsize m}0}-y_{\hbox{\scriptsize M}0}),\label{eqn:insx}\\
	y_{\hbox{\scriptsize m}}(y_{\hbox{\scriptsize m}0};t)&=&y_{\hbox{\scriptsize M}}(t)+\cos(2\epsilon n)(y_{\hbox{\scriptsize m}0}-y_{\hbox{\scriptsize M}0}),\label{eqn:insy}
	\end{eqnarray}
where $x_{\hbox{\scriptsize M}}(t)$ and $y_{\hbox{\scriptsize M}}(t)$ represent the ``reference'' trajectory for the center of the probability distribution corresponding to $y_{\hbox{\scriptsize m}0} = y_{\hbox{\scriptsize M}0} $, and $x_{\hbox{\scriptsize m}}(y_{\hbox{\scriptsize m}0};t)$ and $y_{\hbox{\scriptsize m}}(y_{\hbox{\scriptsize m}0};t)$  stand for the positions of the two classical particles at time $t$ provided that the heavy particle was at $y_{\hbox{\scriptsize m}0}$ at $t=0$. 
The above equations reveal that the distances $x_{\hbox{\scriptsize m}}(y_{\hbox{\scriptsize m}0};t)-x_{\hbox{\scriptsize M}}(t)$ and $y_{\hbox{\scriptsize m}}(y_{\hbox{\scriptsize m}0};t)-y_{\hbox{\scriptsize M}}(t)$ are proportional to the initial distance $y_{\hbox{\scriptsize m}}(y_{\hbox{\scriptsize m}0};0)-y_{\hbox{\scriptsize M}}(0)=y_{\hbox{\scriptsize m}0}-y_{\hbox{\scriptsize M}0}$ and the scaling factors do not depend on $y_{\hbox{\scriptsize m}0}$; in particular,
\begin{equation}
x_{\hbox{\scriptsize m}}(y_{\hbox{\scriptsize m}0};t)-
x_{\hbox{\scriptsize M}}(t)=\frac{\tan(2\epsilon n)}{\epsilon}[y_{\hbox{\scriptsize m}}(y_{\hbox{\scriptsize m}0};t)-y_{\hbox{\scriptsize M}}(t)].
\label{scaling}
\end{equation}

Given the initial distribution $P_{\hbox{\scriptsize xy},0}(x_{\hbox{\scriptsize m}},y_{\hbox{\scriptsize m}};0)$ and the scaling relations \eqref{eqn:insy},\eqref{scaling}, we obtain the  distribution at later times as 
	\begin{eqnarray}\label{eqn:rhs}
	P_{\hbox{\scriptsize xy},n}&&(x_{\hbox{\scriptsize m}},y_{\hbox{\scriptsize m}};t)=N_n\exp\left\{-\frac{\left[y_{\hbox{\scriptsize m}}-y_{\hbox{\scriptsize M}}(t)\right]^2}{2\Delta\sigma_{y,n}^2}\right\}\\
	&&\times\delta\left(x_{\hbox{\scriptsize m}}-
x_{\hbox{\scriptsize M}}(t)-\frac{\tan(2\epsilon n)}{\epsilon}[y_{\hbox{\scriptsize m}}-y_{\hbox{\scriptsize M}}(t)]\right),\nonumber
	\end{eqnarray}
where $N_n$ are normalization factors and 
\begin{equation}
\Delta\sigma_{y,n}=\nolinebreak\Delta\sigma_{y,0}|\cos(2\epsilon n)|
\label{sig-y0}
\end{equation}
[cf. Eq.\eqref{eqn:insy}]. The corresponding marginal distribution $P_{\hbox{\scriptsize y},n}(y_{\hbox{\scriptsize m}};t)$ is, up to a prefactor, equal to the exponential factor in Eq.\eqref{eqn:rhs}, while the marginal distribution with respect to $x_{\hbox{\scriptsize m}}$ is
	\begin{equation}
	P_{\hbox{\scriptsize x},n}(x_{\hbox{\scriptsize m}}) \cong  \exp\left\{-\frac{[x_{\hbox{\scriptsize m}}-x_{\hbox{\scriptsize M}}(t)]^2}{2\Delta\sigma_{x,n}^2}\right\},
	\end{equation}
where $\Delta\sigma_{x,n}=\Delta\sigma_{y,0}|\sin(2\epsilon n)|/\epsilon$ [cf. Eq.\eqref{eqn:insx}].

According to Eqs.(\ref{eqn:rhs}) and (\ref{sig-y0}), the marginal distribution $P_{\hbox{\scriptsize y},n}(y_{\hbox{\scriptsize m}};t)$ contracts to a $\delta$-function when the number of collisions reaches the critical value $n_{\hbox{\scriptsize cr}}\approx\frac{\pi}{4\epsilon}$. Since this value is equal to $n_{\hbox{\scriptsize max}}$, the collisions stop at the same time.

According to Eq.\eqref{eqn:nunhab}, there is a direct connection between the entanglement properties of the wave function of the system in terms of variables $x$ and $y$ and the factorizability of the classical probability distribution   $P_{\hbox{\scriptsize xy},n}(x_{\hbox{\scriptsize m}},y_{\hbox{\scriptsize m}};t)$  in terms of variables $x_{\hbox{\scriptsize m}}$ and $y_{\hbox{\scriptsize m}}$. The former is entangled, if and only if the latter is not factorizable. For example, the initial wave function $\psi_0(x,y)$ is not entangled and the corresponding probability distribution $P_{\hbox{\scriptsize xy},0}(x_{\hbox{\scriptsize m}},y_{\hbox{\scriptsize m}};0)$ given by Eq.\eqref{eqn:initcond} is factorizable. According to Eq.\eqref{eqn:rhs}, the distribution $P_{\hbox{\scriptsize xy},n}(x_{\hbox{\scriptsize m}},y_{\hbox{\scriptsize m}};t)$ becomes not factorizable at later times due to the presence of $y_{\hbox{\scriptsize m}}$ in the $\delta$-function. Therefore, the wave function of the system becomes entangled. However, after the critical number of collisions $n_{\hbox{\scriptsize cr}}$, the Gaussian in Eq.\eqref{eqn:rhs} contracts to the $\delta$-function and as a result $P_{\hbox{\scriptsize xy},n}(x_{\hbox{\scriptsize m}},y_{\hbox{\scriptsize m}};t)$ becomes factorizable again. Thus, the two particles are no longer entangled at the time when the collisions stop.

The narrow-wave-packet assumption~(\ref{ballistic}) remains valid until the last collision if $(\epsilon m_x\sigma_{0x}v_{x0})/(\pi\hbar) \gg 1$\cite{elsupp}.

Substituting Eq.\eqref{eqn:rhs} into Eq.\eqref{eqn:nunhab}, we obtain 
	\begin{equation}
	\psi_n(x,y;t)\sim\exp\Big\{A^{xx}_nx^2\!+\!A^{yy}_ny^2\!+\!A^{xy}_nxy\!+\!B^x_nx\!+\!B^y_ny\Big\},\nonumber
	\end{equation}
where of particular interest is the coefficient
	\begin{eqnarray}
	A^{xy}_n&=&\frac{\sin(4\epsilon n)[\beta_y^2(t)-\epsilon^2\beta_x^2(t)]}{2\sqrt{\epsilon}\beta_x^2(t)\beta_y^2(t)},\label{eqn:result3}
	\end{eqnarray}
which controls the entanglement between the two particles. As anticipated, it is equal to zero for \mbox{$n=0$} and \mbox{$n=n_{\hbox{\scriptsize cr}}$}. The other two $A$-coefficients are 
\mbox{$A^{xx}_n=-\frac{\cos^2(2\epsilon n)}{2\beta_x^2(t)}-\frac{\sin^2(2\epsilon n)}{2\beta_y^2(t)}\epsilon^2$} and \mbox{$A^{yy}_n=-\frac{\cos^2(2\epsilon n)}{2\beta_y^2(t)}-\frac{\sin^2(2\epsilon n)}{2\beta_x^2(t)}\frac{1}{\epsilon^2}$}. We omit the explicit expressions for the coefficients $B^x_n$ and $B^y_n$.

At \mbox{$n=n_{\hbox{\scriptsize cr}}$}, we obtain \mbox{${A^{xx}_{n_{\hbox{\scriptsize cr}}}}=-\frac{\epsilon^2}{2\beta_y^2(t)}=A_0^{yy}\epsilon^2$} and \mbox{${A^{yy}_{n_{\hbox{\scriptsize cr}}}}=-\frac{1}{2\epsilon^2\beta_x^2(t)}=A_0^{xx}\frac{1}{\epsilon^2},$} which implies that, by the time when the collisions stop, the two wave packets exchange the kinetic energies stored in the momentum distributions in their respective rest frames.

We finally mention, that the same results can also be obtained by representing both, the reflection from the wall and the two-particle collision as linear transformations of the coefficients $A^{xx}_n$, $A^{yy}_n$, $A^{xy}_n$, $B^x_n$, $B^y_n$ and then multiplying the transformation matrices. 

In conclusion, we have described a new method for dealing analytically with interacting one-dimensional wave packets and applied it to the system depicted in Fig.~\ref{pic:final}. The essence of the new method is a reduction of a quantum problem to classical calculations. The idea of this method should be generalizable to a broader class of problems with non-Gaussian shapes of the wave packets, larger numbers of particles and different forms of the interaction potential.  The unexpected result is that the entanglement between the two particles, that arises due to the collisions, vanishes by the time when the collisions stop.

\clearpage

\begin{center}
{\bf SUPPLEMENTARY MATERIAL }
\end{center}

\setcounter{figure}{0}
\renewcommand{\thefigure}{S\arabic{figure}}

\setcounter{equation}{0}
\renewcommand{\theequation}{S\arabic{equation}}

Note: In this supplementary material, we use the same notations as in the main article unless explicitly specified otherwise.

\subsection{Full expression for the initial wave function in Eq.(2) of the main article}\label{sec:one}

The sign ``$\sim$'' in Eq.(2) of the main article means that the wave function needs to be slightly modified to exclude the mutual penetration of the two particles and the wall and then normalized. Since the tails are exponentially small, the details of this modification are not important. One just needs to introduce a cut-off factor, which varies sufficiently slowly as a function of $x$ and $y$. A possible way of doing this is
	\begin{eqnarray}\label{cutoff}
	\psi^{(c)}(x,y)&=&N\exp\!\left\{-\frac{[x-x_{\hbox{\scriptsize M}0}]^2}{2\sigma_{0x}^2}+ip_{x0}x-\frac{[y-y_{\hbox{\scriptsize M}0}]^2}{2\sigma_{0y}^2}\!\right\}\nonumber\\
	&&\times\sin\left(\frac{\pi x}{y}\right)\Theta(x)\Theta(y-x),
	\end{eqnarray}
where $N$ is a normalization constant, $\Theta(x)\Theta(y-x)$ are the Heaviside step functions. The sine-factor in Eq.(\ref{cutoff}) preserves the continuity of the wave function at $x=0$ and $x=y$.

Once the light particle approaches either the wall or the heavy particle, the behavior of the respective tail will evolve to follow the result of the ``antisymmetrized image'' procedure described in the next two sections.

\subsection{Reflection of a wave packet from the wall}\label{chp:oneparticle}

Here, we  briefly summarize the ``antisymmetrized image'' method for calculating the reflection of a quantum wave packet from a hard wall. 

The Hamiltonian of a quantum system consisting of one particle and an impenetrable wall is
	\begin{equation}\label{eqn:thistoreduce}
	\hat{H}_x=-\frac{\hbar^2}{2m_x}\frac{\partial^2}{\partial x^2}+S\cdot\Theta(-x)
	\end{equation}
in the limit $S\rightarrow\infty$. In the region $x>0$, the particle is free to move, but it cannot penetrate the region $x\leq0$. Therefore, its wave function should continuously approach $0$ from the side of positive $x$  and stay $0$ on the negative-$x$ side.

Let us consider a wave function, which is well approximated far from the wall by a Gaussian wave packet
	\begin{equation}\label{eqn:evolve}
	\phi(x;t)=N(t)\exp\left\{-\frac{[x-x_c(t)]^2}{2\beta_x(t)^2}+ip_{x0} x\right\}.
	\end{equation}
In order to respect the boundary condition at the wall, we introduce the mirror image of the above wave packet $\phi(-x;t)$, then consider the antisymmetrized superposition of the original packet and the image, and, finally, take only the positive-$x$ part of this superposition:
\begin{equation}\label{eqn:actpart}
	\phi_a(x;t) =\left[\phi(x;t)-\phi(-x;t)\right]\Theta(x).
	\end{equation}
One can check directly that the above wave function is, in fact, an exact solution of the particle-wall problem. 

According to Eq.\eqref{eqn:evolve}, the two wave packets $\phi(x;t)$ and $\phi(-x;t)$  have opposite group velocities. Initially, they move towards each other, then they meet and interfere at $x=0$ and, finally, separate again. Once they have separated, the main contribution to \eqref{eqn:actpart} comes from the Gaussian $\phi(-x;t)$. This is why the approximation $\phi_a(x;t)\approx - \phi(-x;t)$ is valid for sufficiently long times after the collision.

To summarize, the wave function after the reflection from the wall is \emph{the wave function of the mirror image evolving in free space}. What happens during the collision is easily computable, but we are not interested in it.

\subsection{Collision between two hard-core particles}\label{chp:singlescattering}

A collision between two hard-core particles is described by the Hamiltonian 
	\begin{equation}
	\hat{H}=\frac{\hat{p}_x^2}{2m_x}+\frac{\hat{p}_y^2}{2m_y}+S\cdot\Theta(x-y)\label{eqn:compare}
	\end{equation}
in the limit $S\rightarrow\infty$. 

Before the collision, as long as the particles are sufficiently far from each other, the wave function of the system is well approximated as
	\begin{eqnarray}
	\psi(x,y;t)&\sim&\exp\left\{-\frac{\left[x-x_{\hbox{\scriptsize M}}(t)\right]^2}{2\beta_x(t)^2}+ip_{x0}x\right\}\\
	&&\times\exp\left\{-\frac{\left[y-y_{\hbox{\scriptsize M}}(t)\right]^2}{2\beta_y(t)^2}\right\}.\nonumber
	\end{eqnarray}
We perform the standard transformation to the center-of-mass coordinate
	\begin{equation}
	R=\frac{m_xx+m_yy}{m_x+m_y},\label{eqn:revert1}
	\end{equation}
and the relative coordinate
\begin{equation}
	r=x-y\label{eqn:revert2}.
	\end{equation}
The Hamiltonian in the new coordinates reads \mbox{$\hat{H}=\hat{H}_R+\hat{H}_r$}, where
	\begin{eqnarray}
	\hat{H}_R&=&\frac{\hat{p}_R^2}{2M},\\
	\hat{H}_r&=&\frac{\hat{p}_r^2}{2\mu}+S\cdot\Theta(r)\;\;\text{and}\;\;S\rightarrow\infty
	\end{eqnarray}
with the total mass $M$ and the reduced mass $\mu$, respectively
	\begin{eqnarray}
	M&=&m_x+m_y,\\
	\mu&=&\frac{m_xm_y}{m_x+m_y}.
	\end{eqnarray}
The dynamics of the center-of-mass coordinate $R$ is just the free-particle evolution. At the same time, the Hamiltonian $\hat{H}_r$ is exactly the same as the one in Section~\ref{chp:oneparticle}. The only difference from the one-particle situation is that now two variables are present. By analogy with the particle-wall collision, it can be shown that, in order to obtain the wave function after the collision, the initial wave function should be propagated in time as if the two particles were not interacting and then the sign of $r$ should be changed but not the sign of $R$. This fact makes the result of this procedure non-trivial once we rewrite the wave function after the collision in the ($x$,$y$)-coordinates:
	\begin{eqnarray}\label{eqn:knowntous}
	&&\psi_1(x,y;t)\sim\\
	&&\hspace{0.15cm}\exp\left\{-\frac{(m_y-m_x)^2}{2\beta_y^2(t)(m_x+m_y)^2}y^2-\frac{2m_y^2}{\beta_x^2(t)(m_x+m_y)^2}y^2\right\}\nonumber\\
	&&\times\exp\left\{\frac{2m_yx_c(t)}{\beta_x^2(t)(m_x+m_y)}y-ip_{x0}\frac{2m_y}{m_x+m_y}y\right\}\nonumber\\
	&&\times\exp\left\{-\frac{2m_x(m_y-m_x)}{\beta_y^2(t)(m_x+m_y)^2}xy+\frac{2m_y(m_y-m_x)}{\beta_x^2(t)(m_x+m_y)^2}xy\right\}\nonumber\\
	&&\times\exp\left\{-\frac{2m_x^2}{\beta_y^2(t)(m_x+m_y)^2}x^2-\frac{(m_y-m_x)^2}{2\beta_x^2(t)(m_x+m_y)^2}x^2\right\}\nonumber\\
	&&\times\exp\left\{-\frac{(m_y-m_x)x_c(t)}{\beta_x^2(t)(m_x+m_y)}x+ip_{x0}\frac{m_y-m_x}{m_x+m_y}x-\frac{x_c^2(t)}{2\beta_x^2}\right\},\nonumber
	\end{eqnarray}
where $x_c^2$ is given in the main article.
The momenta after the collision $p_{x0}\frac{2m_y}{m_x+m_y}\;\;\text{and}\;\;p_{x0}\frac{m_y-m_x}{m_x+m_y}$ appearing in Eq.\eqref{eqn:knowntous} follow from the classical energy and momentum conservation.

\subsection{Probability distributions for the classical multiple collisions problem}\label{sec:classical}

As a part of the full quantum solution, we need to solve a classical problem with the following initial conditions: A wall is at the origin. The light particle has the initial momentum $p_{x0}$ and a definite initial position $x_{\hbox{\scriptsize M}0}$, i.e. the initial probability distribution of positions is $P_{\hbox{\scriptsize x},0}(x_{\hbox{\scriptsize m}};0)=\delta(x_{\hbox{\scriptsize m}}-x_{\hbox{\scriptsize M}0})$. The heavy particle has zero initial momentum and the Gaussian distribution of possible initial positions
	\begin{equation}
	P_{\hbox{\scriptsize y},0}(y_{\hbox{\scriptsize m}};0)\cong \exp\left\{-\frac{[y_{\hbox{\scriptsize m}}-y_{\hbox{\scriptsize M}0}]^2}{2\Delta\sigma^2_{y,0}}\right\}.
	\end{equation}
Below we calculate the evolutions of the marginal probability distributions $P_{\hbox{\scriptsize x},n}(x_{\hbox{\scriptsize m}};t)$ and $P_{\hbox{\scriptsize y},n}(y_{\hbox{\scriptsize m}};t)$ as functions of the number of collisions $n$ occurred. In order to do this, we first fix an arbitrary initial position of the heavy particle $y_{\hbox{\scriptsize m}0}$ and solve the problem in this specific case. Then we address the evolution of the probability distributions.

\subsubsection{Solution for a given $y_{\hbox{\scriptsize m}0}$}

At $t=0$, we have the following initial situation: the light particle is at $x_{\hbox{\scriptsize M}0}$ with a momentum $p_{x0}$ directed towards the heavy particle, which is at the position $y_{\hbox{\scriptsize m}0}$. 

Before starting with the quantitative analysis, let us summarize the qualitative expectations. Due to the small but nevertheless non-zero mass ratio $\epsilon=\sqrt{m_x/m_y}\ll1$, the light particle will lose its momentum and accelerate the heavy particle at each collision. This transfer of momentum lasts as long as the light particle is faster than the heavy one. After the maximal number of collisions $n_{\hbox{\scriptsize max}}$, the light particle will become too slow to catch up with the heavy one. Because of the smallness of $\epsilon$ and the conservation of energy, the velocity of the heavy particle after the last collision is much smaller than the initial velocity of the light particle $v_{x0}=p_{x0}/m_x$.

To compute the position of the heavy particle, we introduce the function\linebreak $ n(y_{\hbox{\scriptsize m}0};t) \equiv n$, which represents the number of collisions before time $t$. With such a de\-fi\-ni\-tion,
	\begin{equation}
	y_{\hbox{\scriptsize m}}(y_{\hbox{\scriptsize m}0};t)=y(n)+\left[t-\tilde{t}(n)\right] v_y(n),\label{eqn:hierdef}
	\end{equation}
where $y(n)$ and $\tilde{t}(n)$ denote, respectively, the position and the time of the $n$-th collision and $v_y(n)$ the velocity of the heavy particle after the $n$-th collision. 

The velocities $v_y(n)$ and $v_x(n)$ after $n$ collisions can be obtained using the following iteration relations
	\begin{eqnarray}
	v_x(n+1)&=&\frac{m_y-m_x}{m_x+m_y}v_x(n)-\frac{2m_y}{m_x+m_y}v_y(n),\\
	v_y(n+1)&=&\frac{2m_x}{m_x+m_y}v_x(n)+\frac{m_y-m_x}{m_x+m_y}v_y(n),
	\end{eqnarray}
which lead to 
	\begin{eqnarray}
	v_x(n)&=&v_{x0} \cos(n\varphi),\label{eqn:twoeq1}\\
	v_y(n)&=&v_{x0} \epsilon\sin(n\varphi),\label{eqn:twoeq2}
	\end{eqnarray}
where
	\begin{equation}
	\varphi=\arctan\left(\frac{2\epsilon}{1-\epsilon^2}\right)\approx2\epsilon.\label{eqn:phiapp}
	\end{equation}
The maximal number of collisions can be determined from the condition
	\begin{equation}
	v_y(n_{\hbox{\scriptsize max}})=v_x(n_{\hbox{\scriptsize max}})
	\end{equation}
which gives
	\begin{equation}
	n_{\hbox{\scriptsize max}}=\frac{\arctan\left(\frac{1}{\epsilon}\right)}{\arctan\left(\frac{2\epsilon}{1-\epsilon^2}\right)}\approx\frac{\pi}{4\epsilon}.\label{eqn:nmax}
	\end{equation}

To calculate the positions and the times of the collisions, another set of iteration relations can be obtained:
	\begin{eqnarray}
	y(n+1)&=&\frac{v_x(n)+v_y(n)}{v_x(n)-v_y(n)}  y(n),\label{eqn:iteriert}\\
	\tilde{t}(n+1)-\tilde{t}(n)&=&\frac{2}{v_x(n)-v_y(n)}  y(n)\label{eqn:time}.
	\end{eqnarray}
Expanding expression \eqref{eqn:iteriert} in powers of the small parameter $\epsilon$ up to the second order, we obtain
	\begin{equation}
	y(n+1)\approx\left(1+4n\epsilon^2\right)y(n).\label{eqn:firsappr}
	\end{equation}
Since we expect the motion of the heavy particle to be slow, the following approximation is valid
	\begin{equation}
	\frac{y(n+1)-y(n)}{1}\approx\frac{dy(n)}{dn}\approx4n\epsilon^2  y(n)
	\end{equation}
and the solution to this equation is
	\begin{equation}
	y(n)=y_{\hbox{\scriptsize m}0}\exp(2n^2\epsilon^2).\label{eqn:heavyn}
	\end{equation}
Using Eq.\eqref{eqn:heavyn} and making the same approximation, we obtain from Eq.\eqref{eqn:time}
	\begin{equation}
	 \tilde{t}(n)\approx\frac{2y_{\hbox{\scriptsize m}0}}{v_{x0}}n\left[1-\epsilon^2\left(\frac{4}{3}n^2+n+\frac{1}{3}\right)\right].
	\end{equation}
Important to know is the inverse of this function, which gives us the number of collisions occurred during time $t$
	\begin{equation}
	 n(y_{\hbox{\scriptsize m}0};t)\approx\frac{tv_{x0}}{2y_{\hbox{\scriptsize m}0}}-\frac{tv_{x0}\left(2tv_{x0}+y_{\hbox{\scriptsize m}0}\right)\left(tv_{x0}-2y_{\hbox{\scriptsize m}0}\right)}{12y_{\hbox{\scriptsize m}0}^3}\epsilon^2.\label{eqn:dasauch}
	\end{equation}
The right-hand-side of the above equation is implied to be rounded from below. The first term in Eq.(\ref{eqn:dasauch}) is what we expect from an infinite mass particle located at $y_{\hbox{\scriptsize m}0}$. The minus sign in front of the second term is also expected because a non-zero mass ratio $\epsilon$ means that the same number of collisions takes longer time. 

The position of the heavy particle between the $n$-th and the $n+1$-th collisions is
	\begin{equation}
	y_{\hbox{\scriptsize m}}(y_{\hbox{\scriptsize m}0};t)=y(n)+\left[t-\tilde{t}(n)\right] v_y(n).
\label{eqn:conti}
	\end{equation}
Likewise, the position of the light particle $x_{\hbox{\scriptsize m}}(y_{\hbox{\scriptsize m}0};t)$ between the two collisions can be written as
	\begin{equation}\label{eqn:xm}
	x_{\hbox{\scriptsize m}}(y_{\hbox{\scriptsize m}0};t)=\left|y(n)-\left[t-\tilde{t}(n)\right] v_x(n)\right|.
	\end{equation}

\subsubsection{Evolution of the classical probability distributions}\label{sec:mechoftherm}

After obtaining the results for a given initial position of the heavy particle, the evolutions of the marginal pro\-ba\-bi\-lity distributions $P_{\hbox{\scriptsize x},n}(x_{\hbox{\scriptsize m}};t)$ and $P_{\hbox{\scriptsize y},n}(y_{\hbox{\scriptsize m}};t)$ can be calculated. These probability distributions refer to the moments of time between the collisions, when, given the assumption of the narrow wave packets, the number of the preceding collisions can be assumed to be the same for any value of $y_{\hbox{\scriptsize m0}}$. We now focus on the shapes and the widths of these distributions as functions of the number of collisions $n$.

The evolution of the probability distributions can be divided into two alternating stages: the evolution between the collisions and the change due to the collisions. During the first stage, i.e. for a fixed value of $n$, the probability distributions $P_{\hbox{\scriptsize x},n}(x_{\hbox{\scriptsize m}};t)$ and $P_{\hbox{\scriptsize y},n}(y_{\hbox{\scriptsize m}};t)$  only shift along the $x_{\hbox{\scriptsize m}}$- and $y_{\hbox{\scriptsize m}}$-axes with respective velocities $p_{xn}/m_x$ and $p_{yn}/m_y$. Therefore, the shapes and widths of the distributions only change during the second stage when $n$ changes.

\begin{figure*}[t]
	\centering
	\scalebox{0.3}{\includegraphics{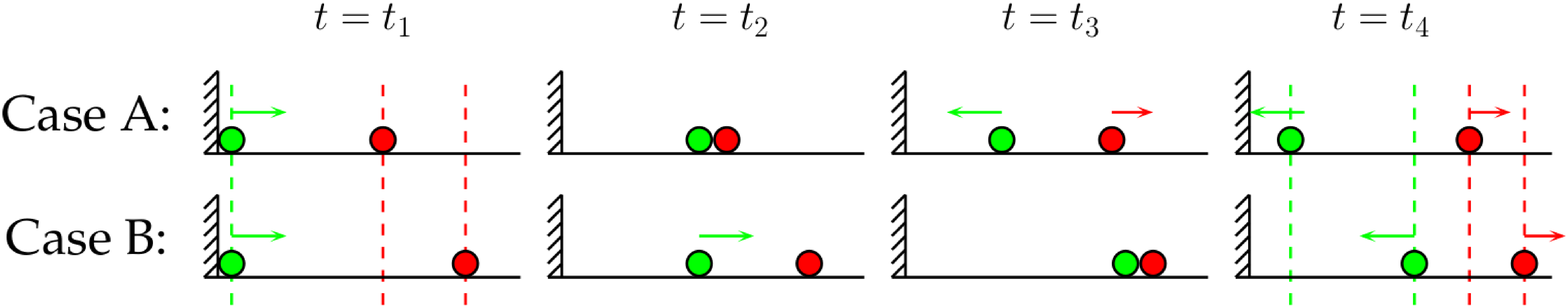}}
	\caption{\small In this sketch, the two rows represent two different initial conditions. The upper and the lower lines of figures represent respectively Cases~A and B described in the text. The columns represent the evolution in time. The arrows depict the velocities of the particles.}
	\label{pic:mechofnarr}
\end{figure*}

As mentioned in the main article, two trends in the evolution of $P_{\hbox{\scriptsize x},n}(x_{\hbox{\scriptsize m}};t)$ and $P_{\hbox{\scriptsize y},n}(y_{\hbox{\scriptsize m}};t)$ can be anticipated qualitatively, namely, an initial broadening of the distribution $P_{\hbox{\scriptsize x},n}(x_{\hbox{\scriptsize m}};t)$ for the light particle and an initial narrowing of the distribution $P_{\hbox{\scriptsize y},n}(y_{\hbox{\scriptsize m}};t)$ for the heavy particle. Both trends can be understood with the help of Fig.~\ref{pic:mechofnarr}, which represents the time evolutions starting from two different initial conditions: Case~A and Case~B.

In the beginning ($t=t_1$), the light particle is at the same position in the both Cases but the heavy particle is not. This is why at a later time ($t=t_2$), the first collision has already happened in Case~A but not yet in Case~B. In Case~B, the particles finally collide at $t=t_3$. After one collision in the both Cases ($t=t_4$), the two velocities of the light particles are the same and the two velocities of the heavy particles are also the same. However, during the time delay between $t_2$ and $t_3$ the heavy particle was moving faster in Case~A. This is why the difference in the heavy particle's positions in the two Cases decreases (distance between the red dashed lines). On the contrary, the distance between the two positions of the light particle has increased during the same interval from zero to a finite value (distance between the green dashed lines).

The changes of the relative distances in Fig.~\ref{pic:mechofnarr} are exaggerated. In reality, these changes are rather small. However, they accumulate after multiple collisions. Moreover, as shown below, the two trends described above additionally enhance each other.

We now calculate the narrowing of the distribution $P_{\hbox{\scriptsize y},n}(y_{\hbox{\scriptsize m}};t)$. For this purpose, we investigate the evolution of the distance between two different initial positions of the heavy particle. Denoting the initial distance as $(\Delta y)_0$ and the distance after $n$ collisions as $(\Delta y)_n$ we obtain
	\begin{equation}
	(\Delta y)_{n-1}-(\Delta y)_n=\Delta v_n \Delta t_n,\label{eqn:this3}
	\end{equation}
where $\Delta v_n$ is the difference in velocity of the heavy particles during the time delay. It follows from Eq.\eqref{eqn:twoeq2}  that
	\begin{equation}\label{eqn:this1}
	\Delta v_n=[\sin(n\varphi)-\sin((n-1)\varphi)]\epsilon v_{x0}.
	\end{equation}
The time delay in Eq.\eqref{eqn:this3} is
	\begin{equation}
	\Delta t_n=\frac{(\Delta x)_{n-1}+(\Delta y)_{n-1}}{[\cos((n-1)\varphi)-\epsilon\sin((n-1)\varphi)]v_{x0}}.\label{eqn:this2}
	\end{equation}
In the numerator, $(\Delta x)_{n-1}$ is the distance between the light particles after $n-1$ collisions. The denominator of Eq.\eqref{eqn:this2} is the difference between the velocities of the heavy and the light particles.

The distance between the light particles can be obtained from
	\begin{equation}
	(\Delta x)_n=(\Delta y)_{n-1}+\Delta t_n  \left[\cos(n\varphi)+\epsilon\sin((n-1)\varphi)\right]v_{x0}.\label{eqn:correctiontodx}
	\end{equation}
In the context of Fig.~\ref{pic:mechofnarr}, the second term in Eq.(\ref{eqn:correctiontodx}) corresponds to the distance the light particle travels in Case~A between $t=t_2$ and $t=t_3$. Inserting \eqref{eqn:this2} into \eqref{eqn:correctiontodx} we obtain
	\begin{equation}
	(\Delta x)_n\approx(\Delta x)_{n-1}+2(\Delta y)_{n-1}\approx\sum_{i=0}^{n-1}2(\Delta y)_i.\label{eqn:approcorrec}
	\end{equation}
Substituting \eqref{eqn:this1} and \eqref{eqn:this2} into \eqref{eqn:this3} and expanding the result in terms of $\epsilon$ up to the second order, we further obtain for a large $n$
	\begin{equation}
	(\Delta y)_{n+1}-(\Delta y)_n\approx-4\epsilon^2\sum_{i=0}^n(\Delta y)_i,\label{eqn:hereapproximation}
	\end{equation}
which, in the continuous limit, can be rewritten as
	\begin{equation}
	\frac{d(\Delta y)_n}{dn}=-4\epsilon^2\int_0^n(\Delta y)_{n'}\ dn'.
	\end{equation}
The solution of this equation is
	\begin{equation}
	(\Delta y)_n=(\Delta y)_0 \cos(2\epsilon n).\label{eqn:advres}
	\end{equation}
The consequence of the dependence (\ref{eqn:advres}) is that the shape of $P_{\hbox{\scriptsize y},n}(y_{\hbox{\scriptsize m}};t)$ remains Gaussian. This originates in the fact that the scaling factor $\cos(2\epsilon n)$ in Eq.\eqref{eqn:advres} does not depend on $y_{\hbox{\scriptsize m}0}$, i.e. the distance between two arbitrary points of the distribution $P_{\hbox{\scriptsize y},n}(y_{\hbox{\scriptsize m}};t)$ becomes rescaled by the same factor as long as all points of the distribution have experienced the same number of collisions. Therefore, it follows from Eq.\eqref{eqn:advres} that the width of $P_{\hbox{\scriptsize y},n}(y_{\hbox{\scriptsize m}};t)$ evolves as
	\begin{equation}
	(\Delta\sigma_y)_n=\Delta\sigma_y |\cos(2\epsilon n)|.\label{eqn:guckda}
	\end{equation}
The explicit form of $P_{\hbox{\scriptsize y},n}(y_{\hbox{\scriptsize m}};t)$ is thus
	\begin{equation}
	P_{\hbox{\scriptsize y},n}(y_{\hbox{\scriptsize m}};t)\cong\exp\left\{-\frac{[y_{\hbox{\scriptsize m}}-y_{\hbox{\scriptsize M}}(t)]^2}{(\Delta\sigma_y)_n}\right\},
	\end{equation}
where $y_{\hbox{\scriptsize M}}(t)=y_{\hbox{\scriptsize m}}(y_{\hbox{\scriptsize M}0};t)$ is given in Eq.\eqref{eqn:conti}.

To calculate $(\Delta x)_n$, we substitute \eqref{eqn:advres} into \eqref{eqn:approcorrec}, which gives
	\begin{equation}\label{eqn:subinto}
	(\Delta x)_n\approx2(\Delta y)_0\sum_{i=0}^{n-1}\cos(2\epsilon i),
	\end{equation}
and then perform the summation with the help of the following approximation:
	\begin{equation}
	\sum_{i=0}^{n-1}\cos(2\epsilon i)\approx\int_0^{n-1}\cos(2\epsilon x)dx=\frac{1}{2\epsilon}\sin(2\epsilon(n-1)).
	\label{eqn:subthis}
	\end{equation}
Substituting Eq.\eqref{eqn:subthis} into \eqref{eqn:subinto} and approximating \mbox{$n-1\approx n$}, we obtain
	\begin{equation}
	(\Delta x)_n\approx\frac{(\Delta y)_0}{\epsilon} \sin(2\epsilon n).\label{eqn:resdsx}
	\end{equation}
Since the scaling factor $\frac{\sin(2\epsilon n)}{\epsilon}$ in Eq.\eqref{eqn:resdsx} does not depend on $y_{\hbox{\scriptsize m}0}$ and $P_{\hbox{\scriptsize y},n}(y_{\hbox{\scriptsize m}};t)$ always has the Gaussian shape, the distribution $P_{\hbox{\scriptsize x},n}(x_{\hbox{\scriptsize m}};t)$ also acquires the Gaussian shape. Its width $(\Delta\sigma_x)_n$ evolves as
	\begin{equation}
	(\Delta\sigma_x)_n=\frac{\Delta\sigma_y}{\epsilon} |\sin(2\epsilon n)|.\label{eqn:guckhier}
	\end{equation}
Equations (12) and (13) of the main article for the evolution of the particles' coordinates follow from Eqs.(\ref{eqn:advres}) and (\ref{eqn:resdsx}) and from the definitions $x_{\hbox{\scriptsize M}}(t)\equiv x_{\hbox{\scriptsize m}}(y_{\hbox{\scriptsize M}0};t)$ [the RHS given by Eq.\eqref{eqn:xm}] and $y_{\hbox{\scriptsize M}}(t) \equiv  y_{\hbox{\scriptsize m}}(y_{\hbox{\scriptsize M}0};t)$ [the RHS given by Eq.\eqref{eqn:conti}].

\subsection{Ballistic spread of the wave packets and the validity of the narrow-wave-packet assumption}\label{sec:ballisticbroad}

The validity of our approximation is limited by the narrow-wave-packet assumption that both $|\beta_x(t)|$ and $|\beta_y(t)|$ should be much smaller than $y_{\hbox{\scriptsize M}}(t)$, which for the sake of an estimate can be further approximated as $y_{\hbox{\scriptsize M0}}$. We limit ourselves to the case $|\beta_x(t)| \gg |\beta_y(t)|$. Therefore, the critical condition invalidating our approximation is  
\begin{equation}
	y_{\hbox{\scriptsize M}0}^2 = |\beta_x(t)|^2=\sigma_{0x}^2\left(1+\frac{t^2\hbar^2}{m_x^2\sigma_{0x}^4}\right)\approx\frac{t^2\hbar^2}{m_x^2\sigma_{0x}^2}.
	\label{condition}
	\end{equation}
Denoting the number of collisions required to reach this condition as $n_{\hbox{\scriptsize bal}}$, then approximating the corresponding time as  	$t\approx 2y_{\hbox{\scriptsize M}0} n_{\hbox{\scriptsize bal}} / v_{x0} $ and, finally, substituting it in Eq.(\ref{condition}), we obtain
	\begin{equation}
	n_{\hbox{\scriptsize bal}} \approx \frac{m_x\sigma_{0x}v_{x0}}{4\hbar}.
\label{eqn:nbal}
	\end{equation}
The condition for the validity of our approximation becomes 
\begin{equation}
n_{\hbox{\scriptsize max}} \ll n_{\hbox{\scriptsize bal}},
\label{cond2}
\end{equation}
where, as derived in the previous section, $n_{\hbox{\scriptsize max}}\approx\frac{\pi}{4\epsilon}$. Condition~(\ref{cond2}) is thus equivalent to
\begin{equation}
	\frac{\epsilon \ m_x\sigma_{0x}v_{x0}}{\pi \hbar} \gg 1.    
\label{cond3}
	\end{equation}

Let us now give a concrete example when the left-hand-side of inequality (\ref{cond3}) is approximately equal to 1. This critical situation may be realized when the light particle is a big molecule with $m_x=10^{-20}$g and the heavy particle is a very small dust grain of mass $m_y=10^{-10}$g. The initial velocity of the molecule is $v_{x0}=10^4\frac{\text{cm}}{\text{s}}$. The initial wave packet widths of the two particles are $\sigma_{0x}=10^{-6}$cm and $\sigma_{0y}=10^{-10}$cm.

In order to satisfy condition (\ref{cond3}) starting from the above numbers, one should increase the value of either of the parameters appearing in the product $\epsilon \  m_x\sigma_{0x}v_{x0}$.

\end{document}